[Original Article]

# Prostate-Specific Foundation Models for Enhanced Detection of Clinically Significant Cancer

Jeong Hoon Lee[1], Cynthia Xinran Li[1], Hassan Jahanandish[1,2], Indrani Bhattacharya[1], Sulaiman Vesal[1,2], Lichun Zhang[1], Shengtian Sang[1], Moon Hyung Choi[3], Simon John Christoph Soerensen[2], Steve Ran Zhou[2], Elijah Richard Sommer[2,4], Richard Fan[2], Pejman Ghanouni[1], Yuze Song[5], Tyler M. Seibert[5,6,7], Geoffrey A. Sonn[1,2,*], Mirabela Rusu[1,2,8,*]

[1] Department of Radiology, Stanford University, Stanford, CA, USA

[2] Department of Urology, Stanford University, Stanford, CA, USA

[3] The Catholic University of Korea, Department of Radiology, College of Medicine, 222 Banpo-daero Seocho-gu, Seoul, Republic of Korea

[4] School of Medicine, Stanford University, Stanford, CA, 94305, USA

[5] Department of Radiation Medicine and Applied Sciences, University of California San Diego, La Jolla, CA, USA

[6] Department of Urology, University of California San Diego, La Jolla, CA, USA

[7] Department of Radiology, University of California San Diego, La Jolla, CA, USA

[8] Department of Biomedical Data Science, Stanford University, Stanford, CA, USA

[*] Equal contribution as corresponding author

**Address correspondence to:**

**Mirabela Rusu, MD**

Stanford University, Department of Radiology, 300 Pasteur, Stanford, 94305, California, USA

E-mail: mirabela.rusu@stanford.edu

**Abstract**

Accurate prostate cancer diagnosis remains challenging. Even when using MRI, radiologists exhibit low specificity and significant inter-observer variability, leading to potential delays or inaccuracies in identifying clinically significant cancers. This leads to numerous unnecessary biopsies and risks of missing clinically significant cancers. Here we present prostate vision contrastive network (ProViCNet), prostate organ-specific vision foundation models for Magnetic Resonance Imaging (MRI) and Trans-Rectal Ultrasound imaging (TRUS) for comprehensive cancer detection. ProViCNet was trained and validated using 4,401 patients across six institutions, as a prostate cancer detection model on radiology images relying on patch-level contrastive learning guided by biopsy confirmed radiologist annotations. ProViCNet demonstrated consistent performance across multiple internal and external validation cohorts with area under the receiver operating curve values ranging from 0.875 to 0.966, significantly outperforming radiologists in the reader study (0.907 versus 0.805, *p*<0.001) for mpMRI, while achieving 0.670 to 0.740 for TRUS. We also integrated ProViCNet with standard PSA to develop a virtual screening test, and we showed that we can maintain the high sensitivity for detecting clinically significant cancers while more than doubling specificity from 15% to 38% (p<0.001), thereby substantially reducing unnecessary biopsies. These findings highlight that ProViCNet's potential for enhancing prostate cancer diagnosis accuracy and reduce unnecessary biopsies, thereby optimizing diagnostic pathways.

# Main

Prostate cancer is one of the most common malignancies in American men and the second leading cause of cancer-related mortality in men in the United States[1,2]. Magnetic resonance imaging (MRI) has emerged as a crucial tool for prostate cancer diagnosis by enabling improved visualization of prostate anatomy and lesions, while ultrasound provides cost-effective imaging guidance for biopsy procedures in real time.[3–5]. With these imaging advancements, MRI-ultrasound fusion biopsy techniques enhance the detection of clinically significant prostate cancers (csPCa) and reduce unnecessary biopsies. This is particularly important given that early detection is associated with excellent 5-year survival rates, often exceeding 98%[6,7]. However, current imaging interpretation is still limited by challenges. In patients undergoing biopsy, radiologists interpreting MRI missed cancer rates of 12% for clinically significant prostate cancer, while in patients undergoing radical prostatectomy [8], 34% of clinically significant and 81% of indolent cancers were missed[9,10]. Additionally, MRI interpretation demonstrates significant inter-observer variability, with specificity reported between 21.9% and 68.5%, depending on diagnostic criteria [11]. These diagnostic limitations can lead to delayed detection and intervention, potentially compromising survival outcomes, as the 5-year survival rates drop to 34% in advanced stages of the disease[12]. To overcome these diagnostic limitations and the associated decline in survival outcomes, accurate interpretation of both MRI and TRUS is essential for enhancing diagnostic precision and guiding appropriate biopsy decision.

Recent developments in artificial intelligence (AI) have demonstrated significant potential in medical image analysis[13]. Particularly, the recent emergence of vision foundation models, which are pre-trained on large-scale datasets and can be adapted to various downstream tasks, has further accelerated the progress in computer vision tasks[14,15]. These vision foundation models have shown improved performance across various domains by learning generalizable representations from vast amounts of diverse medical imaging modalities[16]. They not only serve as powerful feature extractors but also demonstrate enhanced generalization capabilities with limited data and offer robust transfer learning abilities across different medical imaging domains[17,18]. In the field of prostate cancer, while deep learning-based approaches for MRI analysis have shown promising results[9,19,20], there remains a notable absence of foundation models specifically designed for prostate imaging analysis. A specialized foundation model, incorporating prostate-specific anatomical features and imaging characteristics could potentially extend beyond detection. Moreover, it would serve as a versatile tool for various downstream tasks in prostate cancer management, such as screening for biopsy decisions, treatment planning, progression monitoring, and risk stratification.

In this study, we introduce ProViCNet, a model developed to investigate whether vision foundation models could improve the detection and localization of prostate cancer in multi-modal medical imaging including mpMRI and Trans-Rectal Ultrasound (TRUS) (Fig. 1a). Our approach integrates the vision foundation model's general vision capabilities with prostate-specific anatomical knowledge through a specialized training strategy. The framework was designed to process both MRI and ultrasound imaging data, incorporating domain-specific features using label-guided patch-based self-supervised learning while maintaining the advantages of foundation models. This approach aims to enhance diagnostic precision, reduce false positives requiring biopsy confirmation, and decrease inter-observer variability. To evaluate the clinical applicability of our approach, we conducted extensive validation using two internal datasets and external datasets from three independent centers. In addition, we performed comparative analysis against experienced urology specialists. We further evaluated the model's performance against conventional clinical risk stratification methods, including the PI-RADS scoring system and PSA-based screening, for biopsy decision support. This study demonstrates the model's capabilities and its promise in enhancing clinical decision-making for prostate cancer diagnosis, potentially improving patient care and outcomes.

**Results**

*Study cohorts*

We included radiology images from 4,401 patients across six cohorts (Fig. 1c, Table 1), using multi parametric MRI (T2-weighted, Diffusion Weighted Imaging - DWI, and Apparent Diffusion Coefficient -ADC) in all cohorts and additional TRUS imaging for the training dataset, C1, and C4. Of these, 1,404 patients were randomly split 80:20 for training and internal validation (Methods), with model performance evaluated on five cohorts (C1–C5). C1 (n=352) and C2 (n=120) are internal test sets, comprising a biopsy-confirmed cohort and a radical prostatectomy (RP) cohort with pathology-based ground truth, respectively. C3 (n=1497) and C4 (n=1154) are publicly available external datasets [19,21], and C5 (n=292) is an external validation set. Table 1 presents an overview of PSA distributions, Gleason Grade, and lesion characteristics across cohorts. Detailed information about patient selection criteria, clinical characteristics and imaging protocols can be found in the Methods section.

*Architecture of the foundation model*

We developed ProViCNet, a prostate-specific foundation model that integrates MRI and transrectal ultrasound imaging to detect and localize csPCa (Fig. 1a). We performed a lesion-level evaluation, where the prostate was divided into six regions (Fig. 1b, Extended Fig. 1), and area under the receiver operating characteristic curve (AUROC), area under the precision-recall curve (AUPRC), sensitivity, and specificity were calculated using the 90th percentile of prediction values with thresholds determined during internal validation (Methods). ProViCNet employs a 3D-enhanced vision transformer pretrained on the DINOv2 model[15], coupled with patch-level contrastive learning that effectively distinguishes cancer tissue from normal tissue even near ambiguous lesion boundaries (Fig. 1d) (Methods). Each MRI sequence is processed through a dedicated decoder to generate probability maps, which are then fused to capture complementary anatomical and functional details (Extended Data Fig. 1). Additional implementation details, including contrastive pair sampling and model training protocols, are provided in the Methods section.

### *Diagnostic performance of AI model for csPCa*

In the internal biopsy-confirmed test dataset C1, ProViCNet with mpMRI sequences achieved strong discriminative performance with patient-level average AUROC 0.923 and high sensitivity 0.895 while maintaining clinically relevant specificity 0.778 (Fig 2a). The corresponding AUPROC was 0.879, indicating robust performance even with class imbalance. The RP cohort C2, which provided histopathology-derived ground truth labels, achieved AUROC 0.875, AUPROC 0.822, with sensitivity 0.819 and specificity 0.730. The DSC for both C1 and C2 was 0.425 and 0.389, respectively (Fig 2b). Detailed metrics for these cohorts, including sensitivity, specificity, PPV, NPV, DSC, and accuracy, can be found in Extended Data Table 1.

Qualitative analysis of the model predictions revealed the complementary nature of different MRI sequences (Fig. 2d). While T2-weighted images provided detailed anatomical information, DWI and ADC sequences contributed distinctive functional characteristics of the tissue. The integration of these complementary features enabled comprehensive mpMRI predictions. Representative cases with varying segmentation performance are shown in Figure 2e, displaying the axial slice with the largest cancer extent for each case, illustrating the model's performance across different scenarios from low to high Dice scores (0.115-0.603). Metrics for each individual MRI sequence in the C1 and C2 cohort can be found in Extended Data Table 2. For the internal C1 cohort, the AUROC for T2, DWI, and ADC sequences were 0.899, 0.885, and 0.851, respectively. For the C2 cohort, the AUROC values for T2, DWI, and ADC sequences were 0.824, 0.866, and 0.827, respectively, showing slightly lower performance compared to the C1 cohort. This decrease in performance can be attributed to the use of histopathology-derived ground truth labels in the C2 cohort.

For the C1 cohort TRUS data, the AUROC was 0.735, with a sensitivity 0.691, specificity of 0.571, and a Dice score of 0.144 (Figure 2f). For the C4 cohort, the AUROC was 0.670, with sensitivity and specificity of 0.715 and 0.462, respectively, and a Dice score of 0.124 (Extended Data Table 3). Representative examples illustrating segmentation performance variations are shown in Figure 2g, where the axial slice containing the most extensive cancer region per case is depicted. These cases demonstrate the model's performance across different scenarios, with Dice scores ranging from poor (0.000) to strong agreement (0.668).

External validation performance for csPCa

The model was evaluated across multiple external cohorts. In the cohort C3, the model achieved its highest performance with AUROC 0.966 and AUPROC 0.933, along with a sensitivity 0.953 and specificity 0.761. The C4 cohort showed consistent performance with AUROC 0.920 and AUPROC 0.854, maintaining a comparable sensitivity of 0.846 and specificity of 0.766. In the C5 cohort, which had the highest proportion of indolent cancers, the model demonstrated performance of AUROC 0.946, and the highest specificity 0.951 among all cohorts.

*Detection Performance for All Prostate Cancer Lesions*

The model's performance with respect to detecting all prostate cancers, including both indolent and clinically significant cases, is detailed in Extended Table 4. Upon including indolent cancers in the analysis, AUROC values showed modest improvements of 0.5% to 3.2% across cohorts, with the highest increase observed in the C2 cohort (AUROC 0.907 to 0.936). However, this broader detection scope resulted in decreased specificity across all cohorts due to increased false positive predictions, which was particularly notable in C4 where specificity dropped from 0.766 to 0.636, with an average decrease of 9.9% (range: 2.6-13.1%).

*Lesion-level performance*

To evaluate the model's ability to detect individual lesions, we performed a lesion-level analysis where each cancer lesion and each cancer-free sextant was treated as a separate case. The internal biopsy cohort C1 achieved lesion-level AUROC 0.918 (95% CI: 0.894-0.943), while the RP cohort C2 showed AUROC 0.853 (95% CI: 0.813-0.893) (Fig. 2b). In external validation, the PI-CAI cohort C3 demonstrated lesion-level AUROC 0.921 (95% CI: 0.863-0.898), and the UCLA cohort C4 achieved AUROC 0.880 (95% CI: 0.905-0.938).

*Comparative diagnostic performance of the AI and radiologist*

We compared the performance of the AI model with radiologists using a subset of 93

patients who undergo RP with clinically significant cancer from the C2 cohort. Figure 2c summarizes the diagnostic performance of both the AI and radiologists. The AI model demonstrated a significantly higher AUROC of 0.907 compared with the radiologists' 0.805 ($p<0.001$, Wilcoxon test). The AI model showed sensitivity and specificity of 0.880 and 0.654, respectively, while radiologists achieved 0.825 and 0.799. The Dice scores were 0.396 for the AI model and 0.347 for the radiologists (Extended Data Table 5).

*Feature representation analysis*

To evaluate the discriminative capabilities of learned features, we performed visualization analysis on the internal test cohort C1. Features were extracted from the pretrained vision transformer backbone of ProViCNet using small patches from T2-weighted MRI sequences. Up to 10 patches per label category were sampled from each patient's prostate gland. The high-dimensional features were reduced to three components using Uniform Manifold Approximation and Projection (UMAP) for visualization. In the three-dimensional feature space, patches were color-coded according to their tissue labels: background (gray), normal prostate gland (green), indolent cancer (orange), and aggressive cancer (red) (Fig. 3a). Visualization revealed distinct clustering patterns corresponding to different tissue types, suggesting that patch-level representation learning captured discriminative features for distinguishing normal prostate tissue, indolent cancer, and csPCa.

Additionally, we performed component-wise feature visualization Using principal components analysis (PCA) to examine feature patterns across different MRI sequences (Extended Data Fig. 3). The first three PCA components showed distinct spatial patterns for T2-weighted, DWI and ADC sequences within the prostate gland. While individual sequences showed some false positive regions, the final integrated prediction combining all sequences showed reduced false positive signals, particularly in distinguishing csPCa regions from normal prostate tissue.

## *Improving Specificity in PSA-Based Biopsy screening with AI*

Prostate-specific antigen (PSA) is a widely used tool for prostate cancer screening. PSA $\geq 4$ is a widely accepted threshold for recommending biopsy, primarily due to its high sensitivity. However, its low specificity results in a significant number of unnecessary biopsies in patients without csPCa. To address this limitation, we evaluated the ability of mpMRI-based ProViCNet predictions to distinguish patients with csPCa from those without, comparing its performance to PSA $\geq 4$ by analyzing lesion-specific maximum predicted values (Fig. 3b). Across the C1, C3, and C4 cohorts, PSA achieved AUROCs ranging from 0.666 to 0.688. ProViCNet predictions yielded significantly higher AUROCs of 0.843, 0.875, and 0.798, respectively ($p<0.001$, DeLong's test), demonstrating its ability to distinguish between tissue types more effectively.

Next, we evaluated the potential of mpMRI-based ProViCNet predictions to enhance specificity while preserving the sensitivity of PSA $\geq 4$ (Fig. 3c). Across the combined cohorts (C1, C3, and C4), the PSA $\geq 4$ threshold achieved a sensitivity of 0.937 but a specificity of only 0.147. By integrating ProViCNet's AI predictions, specificity improved to 0.378, representing a relative increase of 157%, while sensitivity was maintained. This improvement also increased the overall accuracy from 0.319 to 0.500. These findings highlight the potential for combining MRI-based AI predictions with PSA screening, to reduce the number of unnecessary biopsies without compromising the diagnostic performance.

*Comparison with Existing Segmentation Models*

We conducted a comprehensive comparative analysis between ProViCNet and eight established segmentation models to evaluate the relative performance in prostate cancer detection (Fig. 4a)[22–29]. To ensure standardized comparison, all models were evaluated using only T2-weighted MRI sequences from the C1 cohort, without multi-parametric fusion. Patient-level AUROC evaluation demonstrated that ProViCNet achieved superior performance of AUROC 0.899 compared with other models, with nnUNet showing the second-highest performance of AUROC 0.863. The remaining models achieved AUROC values ranging from 0.848 to 0.710, with particularly notable differences in performance in cases with small lesions and complex anatomical structures (detailed performance metrics in Extended Data Table 6).

Lesion-level performance evaluation using DeLong's test revealed significant differences in AUROC between ProViCNet and nnUNet ($p<0.001$). This performance advantage was consistent across different prostate zones and tumor sizes. Probability heatmap visualization (Fig. 4b) showed the comparison between predicted cancer regions from different models on the same case used in Figure 2d. Evaluation of T2-weighted MRI sequences, revealed that ProViCNet exhibited higher performance in the respect to detecting clinically significant lesions (AUROC 0.899, sensitivity 0.774, specificity 0.874) compared with nnUNet (AUROC 0.863, sensitivity 0.476, specificity 0.974; Extended Data Table 7).

*Morphological correlates of model performance*

Quantitative analysis revealed significant correlations between model performance and morphological characteristics of prostate cancer (Fig 4c-e). The Dice score showed moderately positive correlations with cancer lesion volume (Spearman's $\rho = 0.514$, $p<0.001$) and lesion's

Gleason Grade (ρ = 0.416, *p*<0.001), while prostate volume did not show a significant correlation with prostate volume (ρ = 0.05, *p*=0.620) in the C1 cohort. Analysis of lesion volume quartiles demonstrated a consistent trend across all cohorts, with larger lesions being associated with higher Dice scores. The model's prediction confidence also showed positive correlation with lesion volume (ρ = 0.368, *p*<0.001). This relationship between lesion size and detection accuracy was maintained across different Gleason Grade groups, with particularly robust performance in higher-grade lesions.

*Ablation Study of Model Components*

We performed a systematic ablation study using all mpMRI sequences to evaluate the contribution of each architectural component (Table 2). The baseline ViT architecture achieved an AUROC of 0.747, and showed a performance comparable to that of conventional architectures such as SwinUNet, UNet, and LeViTUNet (Extended Data Table 6)[27–29]. Integration of the DINOv2 pre-trained weights substantially improved model performance (AUROC: 0.877), demonstrating the significant impact of transfer learning from vision foundation models. Alternative approaches, such as using frozen DINOv2 weights with Low-Rank Adaptation (LoRA), showed inferior performance (AUROC: 0.824)[30]. While this performance exceeded that of ViT models trained without pre-trained weights, it suggests that some degree of backbone fine-tuning is necessary for optimal performance on downstream tasks. During fine-tuning, we found that applying a small learning rate (10%) to the backbone yielded optimal model performance. The 3D-enhanced positional embedding tokens further increased the AUROC to 0.918. The final model incorporating patch-level contrastive learning achieved the highest performance (AUROC: 0.930), demonstrating the cumulative benefit of each component.

**Discussion**

Accurate detection and localization of clinically significant prostate cancer remains a critical challenge in clinical practice, impacting millions of men worldwide. In this study, we developed ProViCNet, a prostate-specific vision foundation model that demonstrates robust performance in detecting and localizing prostate cancer across multiple imaging modalities including multi-parametric MRI sequences and TRUS. Our extensive multi-institutional validation confirmed that ProViCNet outperforms both experienced radiologists and conventional AI methods. By combining domain-specific learning strategies with large-scale vision model training, ProViCNet adeptly distinguishes subtle lesion boundaries across imaging modalities, thus offering a promising avenue to enhance prostate cancer diagnosis and reduce inter-observer variability.

ProViCNet addresses several critical challenges in prostate cancer diagnosis. The model's superior performance compared with experienced radiologists (AUROC 0.907 vs 0.805, $p<0.001$), with notably higher sensitivity (0.880 vs 0.825, $p<0.001$) in identifying csPCa, Given the increasing adoption of prostate MRI as a primary diagnostic tool, these performance improvements could be particularly valuable for clinical practice. The model generates lesion probability maps that could aid in biopsy targeting decisions. Additionally, when integrated with PSA screening, ProViCNet offers a practical approach for improving the current diagnostic paradigm. By increasing specificity from 0.147 to 0.378, while maintaining sensitivity 0.937 at PSA $\geq$ 4, the model could potentially reduce unnecessary biopsies by 157% without compromising cancer detection rates. This improvement is especially relevant considering the psychological burden and healthcare costs associated with unnecessary procedures.

The methodological advancements in ProViCNet contribute significantly to its robust performance. While conventional deep learning approaches have shown promise in prostate cancer detection, our approach, leveraging vision foundation model with patch-level representation learning, enables more generalizable feature learning. This strategy proved particularly effective, as demonstrated by our feature representation analysis and ablation studies, enhancing the model's ability to distinguish clinically significant cancers from other prostate tissue. The integration of multi-parametric MRI sequences through sequence-specific decoders allows for comprehensive capture of both the anatomical and functional characteristics of prostate tissue. Interestingly, direct self-supervised learning on prostate imaging did not yield significant improvements in cancer detection performance. We hypothesize that this limitation stems from the characteristics of prostate cancer imaging - the low frequency of cancer regions within images and their diffuse boundaries pose challenges for multi-view contrastive learning approaches like DINOv2, which typically benefit from clear object boundaries. Our findings suggest that while general vision foundation models provide valuable initialization, generating discriminative feature representations for subtle and sparse cancer regions within 3D medical images requires an approach closer to supervision. This is exemplified by our label-guided patch-level contrastive learning strategy, which effectively addresses the challenges of learning from ambiguous cancer boundaries, and low tumor-to-background ratios typical in prostate imaging.

Recent large-scale efforts, including the PI-CAI challenge and specialized frameworks such as FocalNet, CorrSigNIa, and SPCNet, have made substantial progress in mpMRI-based prostate cancer detection [9,19,20,31]. ProViCNet builds upon these advances while exploring a different technical direction through the use of vision foundation model and label-guided patch-level contrastive learning. Through DINOv2-pretrained vision transformer architecture, ProViCNet effectively captures the contextual relationships essential for identifying sparse and indistinct cancer regions in prostate imaging. The 3D-enhanced positional embedding tokens further strengthen the model's ability to learn volumetric structures, while the label-guided patch-level contrastive learning in ViT backbone refines these embeddings, mitigating uncertainties from radiologist-defined lesion borders and enhancing generalization. Collectively, these design elements complement existing approaches by providing more robust feature representations, potentially enabling advanced downstream applications such as treatment outcome, recurrence and survival prediction.

Several limitations of our study should be considered. Our current 2D vision foundation model backbone with 3D positional encoding exhibits strong performance. However, this architecture may not be optimal for truly volumetric imaging modalities like ultrasound and CT, where depth information is as significant as width and height. Although studies suggest minimal performance differences between 2D and 3D approaches in MRI due to large inter-slice distances, future development of native 3D vision transformers could potentially enhance feature extraction from volumetric data. Nonetheless, the 2D backbone offers the advantage of being easily adaptable to 3D architectures and can be applied to a broader range of tasks[32]. Second, while our study included multiple external validation cohorts, there were notable differences in clinical characteristics, particularly in the proportion of clinically significant cancers, across datasets. Additionally, scanner manufacturers varied significantly between cohorts - our internal cohort predominantly used GE scanners (84.55%; Extended data table 8), while external datasets, PI-CAI, were acquired exclusively using Siemens and Philips Medical Systems. This difference could be attributed to several factors, not only the manufacturer, but also including PI-CAI's substantially larger training dataset from their cohorts and different evaluation methodologies. Despite these variations in scanner manufacturers, patient characteristics, and evaluation approaches, our model's robust performance across multiple cohorts demonstrates its potential generalizability in real-world clinical settings. Additionally, although we demonstrated improved performance compared with conventional methods and radiologist interpretation, prospective clinical trials are needed to validate whether these improvements translate to better patient outcomes. In particular, our model provides probability heatmaps indicating regions most likely to contain cancer, potentially guiding more precise needle placement during biopsy procedures. Nonetheless, its impact on biopsy yield and clinical decision-making requires further investigation. Future work should include reader studies to quantify how ProViCNet's assistance affects radiologists' detection performance and its potential role in reducing unnecessary biopsies in real-world clinical settings.

In conclusion, ProViCNet represents a significant advancement in imaging analysis for prostate cancer, demonstrating robust performance across multiple validation cohorts and imaging modalities. The model's ability to process multi-parametric MRI and Ultrasound data and provide interpretable cancer probability maps could enhance diagnostic precision and biopsy guidance. Future work should focus on prospective clinical validation through reader studies to quantify its impact on radiologists' performance and patient outcomes, ultimately establishing its role in improving clinical decision-making for prostate cancer diagnosis.

**References**

1. Siegel, R. L., Giaquinto, A. N. & Jemal, A. Cancer statistics, 2024. *CA Cancer J Clin* **74**, 12–49 (2024).

2. Rawla, P. Epidemiology of prostate cancer. *World J Oncol* **10**, 63 (2019).

3. Kasivisvanathan, V. *et al.* MRI-targeted or standard biopsy for prostate-cancer diagnosis. *New England Journal of Medicine* **378**, 1767–1777 (2018).

4. Ahdoot, M. *et al.* MRI-targeted, systematic, and combined biopsy for prostate cancer diagnosis. *New England Journal of Medicine* **382**, 917–928 (2020).

5. Drost, F.-J. H. *et al.* Prostate MRI, with or without MRI-targeted biopsy, and systematic biopsy for detecting prostate cancer. *Cochrane Database of Systematic Reviews* (2019).

6.  Siegel, D. A. Prostate cancer incidence and survival, by stage and race/ethnicity—United States, 2001–2017. *MMWR Morb Mortal Wkly Rep* **69**, (2020).

7.  Surveillance Epidemiology & Program, E. R. (SEER). Cancer Stat Facts: Prostate Cancer. Preprint at (2024).

8.  Ahmed, H. U. *et al.* Diagnostic accuracy of multi-parametric MRI and TRUS biopsy in prostate cancer (PROMIS): a paired validating confirmatory study. *The Lancet* **389**, 815–822 (2017).

9.  Bhattacharya, I. *et al.* Selective identification and localization of indolent and aggressive prostate cancers via CorrSigNIA: an MRI-pathology correlation and deep learning framework. *Med Image Anal* **75**, 102288 (2022).

10. Johnson, D. C. *et al.* Detection of individual prostate cancer foci via multiparametric magnetic resonance imaging. *Eur Urol* **75**, 712–720 (2019).

11. Simmons, L. A. M. *et al.* The PICTURE study: diagnostic accuracy of multiparametric MRI in men requiring a repeat prostate biopsy. *Br J Cancer* **116**, 1159–1165 (2017).

12. Society, A. C. Survival Rates for Prostate Cancer. Preprint at (2024).

13. Bhattacharya, I. *et al.* A review of artificial intelligence in prostate cancer detection on imaging. *Ther Adv Urol* **14**, 17562872221128792 (2022).

14. Caron, M. *et al.* Emerging properties in self-supervised vision transformers. in *Proceedings of the IEEE/CVF international conference on computer vision* 9650–9660 (2021).

15. Oquab, M. *et al.* Dinov2: Learning robust visual features without supervision. *arXiv preprint arXiv:2304.07193* (2023).

16. Krishnan, R., Rajpurkar, P. & Topol, E. J. Self-supervised learning in medicine and healthcare. *Nat Biomed Eng* **6**, 1346–1352 (2022).

17. Pai, S. *et al.* Foundation model for cancer imaging biomarkers. *Nat Mach Intell* **6**, 354–367 (2024).

18. Zhang, S. & Metaxas, D. On the challenges and perspectives of foundation models for medical image analysis. *Med Image Anal* **91**, 102996 (2024).

19. Saha, A. *et al.* Artificial intelligence and radiologists in prostate cancer detection on MRI (PI-CAI): an international, paired, non-inferiority, confirmatory study. *Lancet Oncol* (2024).

20. Seetharaman, A. *et al.* Automated detection of aggressive and indolent prostate cancer on magnetic resonance imaging. *Med Phys* **48**, 2960–2972 (2021).

21. Natarajan, S., Priester, A., Margolis, D., Huang, J. & Marks, L. Prostate MRI and Ultrasound With Pathology and Coordinates of Tracked Biopsy (Prostate-MRI-US-Biopsy) (version 2). Preprint at https://doi.org/10.7937/TCIA.2020.A61IOC1A (2020).

22. Isensee, F., Jaeger, P. F., Kohl, S. A. A., Petersen, J. & Maier-Hein, K. H. nnU-Net: a self-configuring method for deep learning-based biomedical image segmentation. *Nat Methods* **18**, 203–211 (2021).

23. Wang, H., Cao, P., Wang, J. & Zaiane, O. R. Uctransnet: rethinking the skip connections in u-net from a channel-wise perspective with transformer. in *Proceedings of the AAAI conference on artificial intelligence* vol. 36 2441–2449 (2022).

24. Zhou, Z., Siddiquee, M. M. R., Tajbakhsh, N. & Liang, J. Unet++: Redesigning skip connections to exploit multiscale features in image segmentation. *IEEE Trans Med Imaging* **39**, 1856–1867 (2019).

25. Huang, X., Deng, Z., Li, D. & Yuan, X. Missformer: An effective medical image segmentation transformer. *arXiv preprint arXiv:2109.07162* (2021).

26. Chen, J. *et al.* Transunet: Transformers make strong encoders for medical image segmentation. *arXiv preprint arXiv:2102.04306* (2021).

27. Xu, G., Zhang, X., He, X. & Wu, X. Levit-unet: Make faster encoders with transformer for medical image segmentation. in *Chinese Conference on Pattern Recognition and Computer Vision (PRCV)* 42–53 (2023).

28. Cao, H. *et al.* Swin-unet: Unet-like pure transformer for medical image segmentation. in *European conference on computer vision* 205–218 (2022).

29. Ronneberger, O., Fischer, P. & Brox, T. U-net: Convolutional networks for biomedical image segmentation. in *Medical image computing and computer-assisted intervention–MICCAI 2015: 18th international conference, Munich, Germany, October 5-9, 2015, proceedings, part III 18* 234–241 (2015).

30. Hu, E. J. *et al.* Lora: Low-rank adaptation of large language models. *arXiv preprint arXiv:2106.09685* (2021).

31. Cao, R. *et al.* Joint prostate cancer detection and Gleason score prediction in mp-MRI via FocalNet. *IEEE Trans Med Imaging* **38**, 2496–2506 (2019).

32. Jiao, J. *et al.* USFM: A universal ultrasound foundation model generalized to tasks and organs towards label efficient image analysis. *Med Image Anal* **96**, 103202 (2024).

33. McInnes, L., Healy, J. & Melville, J. Umap: Uniform manifold approximation and projection for dimension reduction. *arXiv preprint arXiv:1802.03426* (2018).

**Methods**

***Study design and datasets***

This retrospective multi-center study was approved by the Institutional Review Board at Stanford University (Protocol: IRB-44998) which waived the requirement for written informed consent. We analyzed multi-parametric MRI data from 1,876 patients to develop and internally validate our model (Fig. 1c, Table 1). The development dataset (1,404 scans from 1404 patients) was randomly split into training (80%) and internal validation (20%) sets. Model performance was evaluated on two internal test cohorts: a biopsy-confirmed cohort (C1, 352 scans from 352 patients) and a radical prostatectomy (RP) specimen cohort (C2, 120 scans from 120 patients). Ground truth labels for cancer and the prostate gland were derived from biopsy-confirmed radiologist annotations for the development and C1 cohorts while for the C2 cohort, we utilized more precise labels derived from AI-detected cancer cell locations in registered H&E histopathology slides.

For external validation, we used two public datasets and one independent institutional cohort. The public datasets included the Prostate Imaging: Cancer AI (PI-CAI) challenge dataset (C3, 1,497 scans from 1,473 patients) and the UCLA prostate cancer dataset (C4, 1,154 scans from 760 patients) [19,21]. Additional external validation was performed using data from UCSD (C5, 292 scans from 292 patients). These external cohorts, representing different institutions and geographical regions, provided diverse patient populations and imaging protocols to evaluate model generalizability.

*Baseline characteristics of patients and image datasets*

For internal cohorts, MRI-ultrasound fusion guided biopsy was performed using the Artemis System (Eigen Health, Grass Valley, California), equipped with Hitachi Ultrasound Devices. Suspicious lesions, primarily with PI-RADS scores ≥ 3, were targeted and projected onto ultrasound images via the fusion system. Ground truth labels were derived from biopsy-confirmed radiologist annotations for most cohorts, while the RP cohort (C2) utilized AI-detected cancer cell locations from registered H&E histopathology slides.

Patient characteristics showed distinct patterns across cohorts (Table 1). The average of PSA levels ranged from 8.9±10.5 ng/mL in the C4 cohort to 11.9±15.0 ng/mL in the C3 dataset. The distribution of maximum ISUP Grade Group (GG) groups revealed distinct population characteristics: C2 which consists of the RP patients showed the highest proportion of GG≥2 (80.0%), reflecting the selection bias inherent in surgical candidates who typically have more aggressive disease. Meanwhile the C4 cohort closely matched our development dataset (41.7% vs 40.7% GG≥2). In contrast, both C3 and C5 cohorts demonstrated notably lower frequencies of GG≥2 (14.6% and 14.7%, respectively). Consistent with their lower-grade disease profile, these cohorts also exhibited smaller mean cancer sizes (1.51±7.0 mm and 1.1±4.8 mm, respectively). The apparently smaller cancer size in the C2 cohort (1.2±3.4 mm) reflects the pixel-level precision of histopathology-derived annotations rather than true biological differences.

*Image acquisition and preprocessing*

Multi-parametric MRI protocols included T2-weighted, diffusion-weighted imaging (DWI), and apparent diffusion coefficient (ADC) sequences. For our internal cohorts for model development and test, MRI examinations were predominantly (84.55%) performed on 3T scanners (GE Healthcare, Chicago, IL) with an endorectal coil (Extended Data Table 8). MRI-ultrasound fusion-guided biopsy was performed using the Artemis System (Eigen Health, Grass Valley, California), with Hitachi Ultrasound Devices. Transrectal ultrasound (TRUS) images were acquired at a frequency of 7.5-10 MHz using 2D end-fire probes, that was reconstructed in 3D and projected on a grid with uniform voxel spacing of $0.5 \times 0.5 \times 0.5$ mm.

For standardization across institutions, T2-weighted MRI sequences served as the reference for spatial normalization. All MRI sequences were resampled to a standardized voxel spacing of $3.0 \times 0.5 \times 0.5$ mm in the axial, coronal, and sagittal dimensions respectively. Images were center-cropped to $256 \times 256$ pixels in the axial plane. Image intensities were normalized using mean and standard deviation calculated from prostate-specific regions. The model input consisted of three consecutive axial slices to incorporate volumetric information while maintaining computational efficiency. All image processing was performed using the SimpleITK library on NIfTI format (nii.gz) data.

*ProViCNet architecture*

ProViCNet is a prostate-specific foundation model designed to detect cancer locations and distinguish csPCa from multi-parametric MRI sequences of prostate cancer patients (Fig. 1a). It consists of several key components (Fig. 1d), (1) input data preprocessing, which utilizes three consecutive axial slices to maintain spatial context; (2) a ViT backbone pretrained using the vision foundation model DINOv2[15]; and (3) weakly supervised learning with patch-level contrastive learning guided by radiologist annotations. The 3D-enhanced vision transformer incorporates the relative axial positions of consecutive slices through positional embedding, creating a unified token representation that preserves spatial relationships. The patch-level contrastive learning strategy enhances the model's ability to differentiate cancer tissue from normal tissue through careful pair selection. Positive pairs are created from patches sharing similar characteristics (either both cancer or both normal tissue patches, with ≥95% overlap), while negative pairs contrast cancer patches with normal tissue patches. To account for potential annotation uncertainties at cancer boundaries, normal patches in negative pairs are sampled at least one patch width away from cancer regions. This mitigates the impact of inherent limitations in radiologist-defined boundaries. This approach also creates robust feature representations that are less sensitive to annotation ambiguities while maintaining strong discriminative power. The model processes three MRI sequences (T2-weighted, DWI, and ADC) independently through sequence-specific decoders that generate pixel-level probability maps for prostate gland segmentation and cancer classification, distinguishing between indolent cancer and csPCa, while TRUS images are processed through a single-sequence decoder. These MRI features are ultimately integrated through a multi-parametric fusion module to enable comprehensive detection and assessment of the prostate gland, cancer, and csPCa (Extended Data Figure 2).

*Patch-level contrastive learning*

We implemented a patch-level contrastive strategy to reinforce the model's discriminative capacity while accommodating uncertainties in radiologist-defined lesion boundaries (Fig. 1d). Each token from the ViT final feature map corresponds to a 14×14 pixel region in the original image, and we superimpose the ground-truth label to determine whether that token is dominantly cancerous or non-cancerous. Specifically, for a patch ρ we define

$$\rho_c(p) = \frac{N_c(p)}{N}, \quad \rho_g(p) = \frac{N_g(p)}{N}$$

where $N_c(p)$ and $N_g(p)$ represent the numbers of cancer and prostate-gland pixels (respectively) within patch $p$ and $N$ is the total number of pixels in that patch. A patch is classified as cancer if $\rho_c(p) \geq 0.95$ or normal gland if $\rho_g(p) \geq 0.95$.

During training, anchor patches are drawn from strongly cancer-positive areas (i.e., high $\rho_c$), then positive pairs are formed with spatially adjacent patches that share a similarly high $\rho_c$. Meanwhile, negative pairs are formed by comparing anchor patches to patches with a minimal cancer proportion (i.e., $\rho_g(p) \geq 0.95$) located at least one patch-distance away. This ensures that boundary regions—where labeling may be uncertain—are excluded from negative pairs, thereby reducing mislabeled examples.

Let $f_a$ and $\boldsymbol{f_t}$ be the feature embeddings (extracted by a contrastive projection head) for an anchor patch $p_a$ and its target patch $\boldsymbol{p_t}$. We compute the cosine similarity $s$ as

$$s(f_a, f_t) = \frac{f_a \cdot f_t}{|f_a|\,|f_t|},$$

and the patch-level contrastive loss $L_c$ follows:

$$L_c = \begin{Bmatrix} 1 - s(f_a, f_t), & \text{(positive pair)} \\ \max(0, s(f_a, f_t) - m) & \text{(negative pair)} \end{Bmatrix}$$

where m=0.5m=0.5m=0.5 is a margin threshold forcing negative patches to remain sufficiently dissimilar. This formulation drives adjacent cancer patches closer in feature space while pushing clearly non-cancer patches farther away.

To further encourage high-fidelity embeddings, we employ a projection head that expands each 384-dimensional ViT token embedding into a higher-dimensional vector with 65,536 dimensions via:

$$h = \text{MLP}(x), \quad h' = \frac{h}{|h|_2}, \quad z = W h'$$

where $x$ is the ViT output token, "MLP" is a three-layer perceptron with batch-normalization and GELU activation, and $W$ is a weight-normalized linear transformation. This high-dimensional projection fosters fine-grained discrimination between subtle normal–cancer differences, while the normalization layers stabilize training. The patch-level contrastive loss is then combined with standard segmentation loss (e.g., cross-entropy or Dice) in an end-to-end fashion:

$$L_{\text{final}} = (1 - \alpha) L_{\text{seg}} + \alpha L_{\text{contrastive}}$$

where α controls the trade-off between the losses. This hybrid objective ensures robust localization of lesions (via segmentation) while learning more discriminative features for cancerous vs. non-cancerous tissue under imperfect boundary annotations. Full implementation details, including code for sampling patch pairs and under-sampling negative vs. positive pairs to avoid class imbalance, are provided in the Supplementary Methods.

*Feature visualization and analysis*

To visualize learned feature representations, we employed both global and local visualization strategies. Patch-level features were spatially interpolated from the model's native patch resolution to match the original image dimensions, enabling direct comparison with input images. For analyzing the distribution of features across different tissue types, we extracted patches from all patients and applied Uniform Manifold Approximation and Projection (UMAP) for dimensionality reduction. This allows visualization of the relationships between normal tissue, indolent cancer, and aggressive cancer patches in a common embedding space[33].

To examine feature distributions specifically within prostate tissue, we focused on patches contained entirely within the prostate gland. These features were analyzed using Principal Component Analysis (PCA), with the top three components visualized using a jet colormap to highlight spatial patterns of learned features. This approach revealed distinct organizational patterns of features between normal and cancerous regions while maintaining anatomical context.

*Model training and optimization*

We employed a multi-task learning strategy with carefully controlled learning dynamics to balance feature adaptation and cancer detection. The ViT backbone, initialized with DINOv2 pre-trained weights, was fine-tuned at a reduced learning rate (10% of the base rate) to preserve the foundational visual features while allowing adaptation to prostate-specific characteristics. This differential learning rate strategy proved crucial for maintaining the generalization capabilities of the vision foundation model while enabling domain-specific optimization.

The model was trained end-to-end using the Adam optimizer with an initial base learning rate of 2 × 10-4 and weight decay of 1 × 10-5. The loss function combined segmentation loss with patch-level contrastive learning loss at a ratio of 9:1, allowing the model to focus primarily on accurate cancer detection while benefiting from the improved feature representations induced by contrastive learning. Each training batch consisted of 32 sets of three consecutive axial slices, with training continued until convergence, typically requiring 100 epochs. The best model was selected based on validation set performance using the average patient-level AUROC from lesion-level evaluation from internal validation dataset. The model was implemented in PyTorch (version 2.0.0), and trained on a server equipped with eight NVIDIA A100 GPUs, each with 48 GB of memory. Training continued until convergence, typically requiring 100 epochs, with the best model selected based on validation set performance.

*Evaluation metrics*

Model performance was evaluated using both lesion-level and pixel-level metrics [9]. For lesion-level evaluation, the prostate was segmented into six distinct regions (sextants) (Fig. 1b). Each region without cancer was labeled as a negative lesion. For regions with cancer, only the cancerous area was assigned a positive lesion label (Extended Fig 1a). To assess predictive performance, we used the 90th percentile of prediction values from each lesion label's pixels, to calculate the AUROC for lesion detection (Extended Fig 1b). The model's predictions were assessed by calculating the AUROC and AUPROC, using the 90th percentile of prediction values within each region. Sensitivity, specificity, PPV, and NPV were also calculated, using the best threshold determined from internal validation on the developmental set.

For patient-level analysis, performance metrics included sensitivity, specificity, PPV, and NPV, calculated using thresholds determined from the internal validation set. csPCa was defined as Gleason Grade Group ≥2. We also evaluated model performance stratified by lesion volume and Gleason Grade groups to assess the impact of tumor characteristics on detection accuracy. The Dice similarity coefficient (DSC) was used to assess spatial overlap between predicted and ground truth cancer regions.

*Statistical analysis*

Statistical analyses were performed using Python (version 3.8.19). Confidence intervals for AUROC and AUPROC were calculated using DeLong's method. Differences in model performance across cohorts and between the model and radiologists were assessed using two-sided Wilcoxon signed-rank tests. Correlations between model performance and morphological characteristics were evaluated using Spearman's correlation coefficient. P values < 0.05 were considered statistically significant.

*Development of Combined PSA-AI Screening Model*

To improve screening specificity while maintaining the sensitivity of PSA ≥ 4, we developed a stacking ensemble model using logistic regression model to integrate the binary PSA threshold status with AI-derived predictions. The model was formulated as:

$$\text{Outcome} \sim \beta_0 + \beta_1 \, (\text{PSA} \geq 4) \; + \; \beta_2 \, (\text{AI prediction}).$$

where Outcome is a binary variable indicating the presence (1) or absence (0) of clinically significant prostate cancer, PSA ≥ 4 is a binary indicator of PSA threshold status, and AI prediction represents ProViCNet's predicted probability of csPCa. The coefficients $\beta_0$, $\beta_1$, and $\beta_2$ were estimated using logistic regression. This stacking architecture enables bidirectional risk reclassification by leveraging complementary information from both clinical biomarkers and AI-derived imaging features. The ensemble approach allows for (1) identification of high-risk cases with PSA < 4 ng/mL through strong AI predictions, and (2) reclassification of PSA ≥ 4 cases as low-risk based on AI predictions. We calibrated the model's decision threshold to match the sensitivity of conventional PSA ≥ 4 screening and evaluated performance through standard diagnostic metrics including sensitivity, specificity, PPV, NPV, and overall accuracy.

*Code availability*

The complete implementation of ProViCNet (initial commit January 2025) is freely available at https://github.com/pimed/ProViCNet. All analyses were performed using Python version 3.8.19. The deep learning models were developed using PyTorch version 2.0.0, with additional dependencies including SimpleITK for image processing and matplotlib version 3.7.5 for visualization. Detailed documentation, including model architecture specifications, training protocols, and inference procedures, are provided in the github repository README file. The source code is released under the MIT license to encourage broad academic and research use. The pre-trained model weights are accessible through the Hugging Face model repository (https://huggingface.co/pimed/ProViCNet ).

**Data availability**

The PI-CAI (C3) and UCLA (C4) datasets used for external validation are publicly available through their respective data portals. Due to privacy regulations, the internal cohort data from Stanford University and external validation data from UCSD (C5) are not publicly available. However, qualified researchers may request access to the internal dataset for academic purposes through appropriate institutional data sharing agreements. Example data and trained model weights sufficient to reproduce our main findings are available in the code repository.

*Reporting Summary*

Further information on research design is available in the Nature Research Reporting Summary linked to this article and in Extended Data Table 8.

**Acknowledgements**

This work was supported by Stanford University (Departments: Radiology, Urology) and by National Cancer Institute, National Institutes of Health (R37CA260346). The content is solely the responsibility of the authors and does not necessarily represent the official views of the National Institutes of Health. We acknowledge Medicanvas and Seou Kim for providing professional assistance in creating the scientific illustrations for this study.

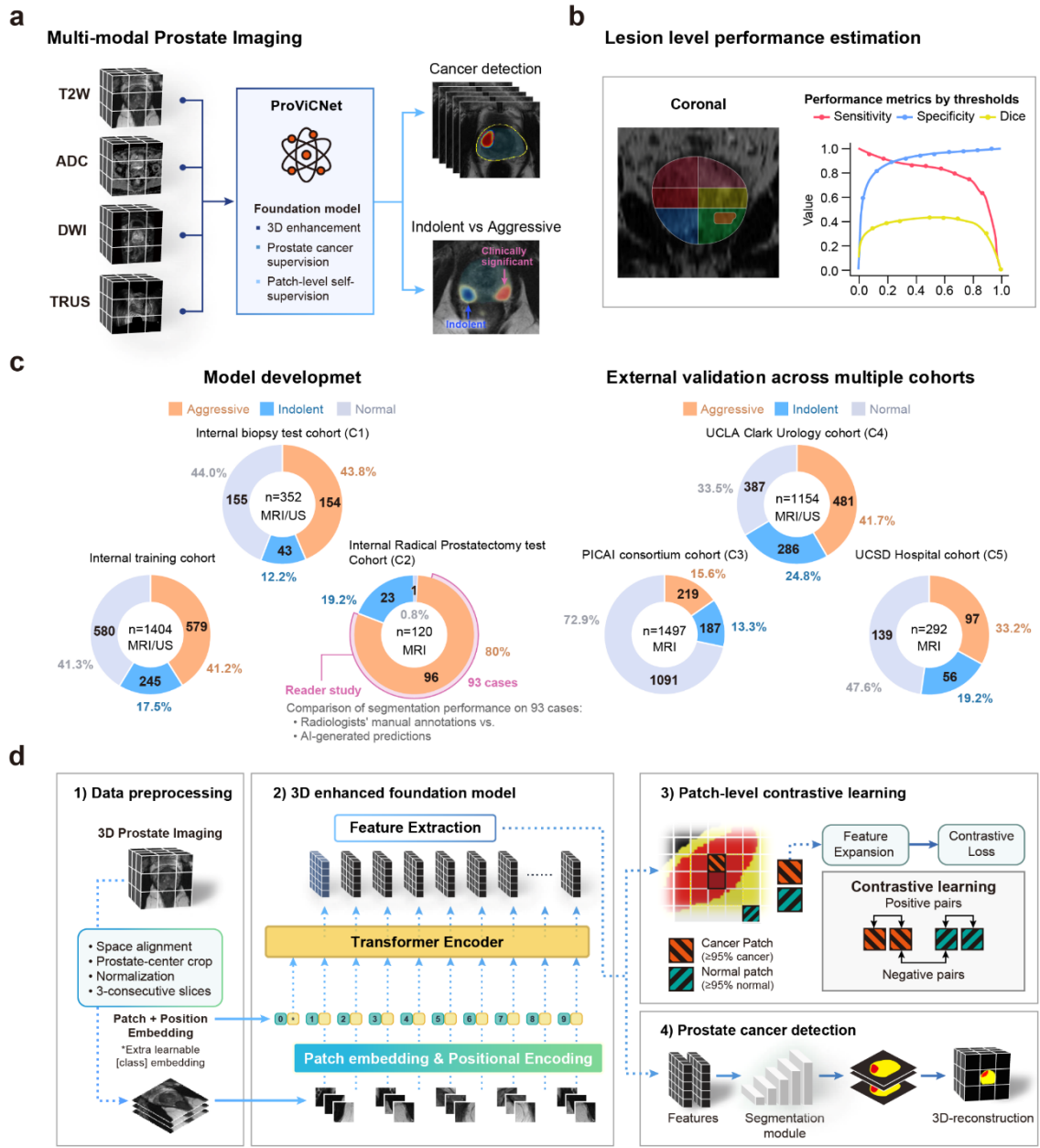

**Fig. 1 | Development and validation of ProViCNet for prostate cancer detection. a**, Overview of the proposed foundation model architecture. ProViCNet processes multi-modal imaging data including multi-parametric MRI sequences (T2W, ADC, DWI) and transrectal ultrasound (TRUS) to detect cancer locations and distinguish between indolent and clinically significant cancers. **b**, Lesion-level performance evaluation framework. The prostate is divided into six regions (sextants), with performance metrics calculated using the 90th percentile of

prediction values for each region. Right: Performance metrics across different classification thresholds. **c**, Study cohorts and population characteristics. Internal training cohort (n=1,404), biopsy-confirmed test (C1, n=352), and radical prostatectomy (C2, n=120) cohorts. External validation was performed on three independent cohorts: PICAI (C3, n=1,497), UCLA (C4, n=1,154), and UCSD (C5, n=292). Pie charts show the distribution of csPCa, indolent cancer, and normal cases. **d**, Technical components of ProViCNet: (1) preprocessing of 3D prostate imaging data, (2) 3D-enhanced foundation model with transformer encoder, (3) patch-level contrastive learning strategy, and (4) final cancer detection module.

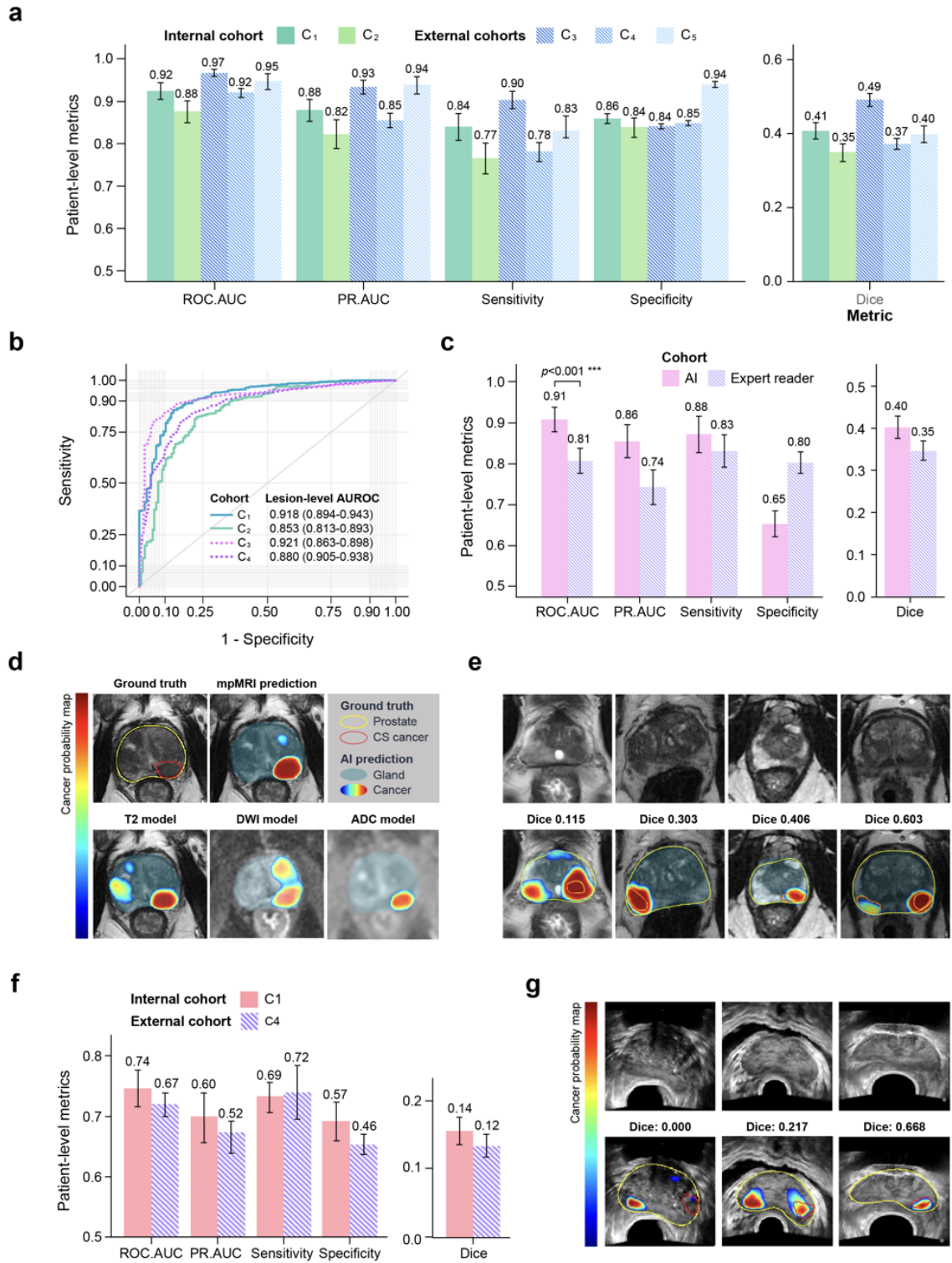

**Fig. 2 | Performance evaluation of ProViCNet across multiple cohorts and comparison with expert readers. a**, Patient-level performance metrics for clinically significant prostate cancer detection across internal (T11, T12) and external (T21, T22, T23) cohorts. Bar plots

show ROC-AUC, PR-AUC, sensitivity, specificity, and Dice similarity coefficient (DSC) with error bars indicating 95% confidence intervals. **b**, Lesion-level receiver operating characteristic curves for four cohorts, with corresponding AUROC values and 95% confidence intervals. **c**, Comparative analysis between ProViCNet and expert radiologists on 93 radical prostatectomy cases, showing statistical significance in ROC-AUC (p<0.001, Wilcoxon test). **d**, Representative case showing multi-parametric MRI prediction integration. Left: ground truth annotation; Right: model predictions from individual MRI sequences (T2-weighted, DWI, and ADC) and their fusion. Color map indicates cancer probability. **e**, Example cases demonstrating varying model performance, arranged by increasing Dice scores (0.115 to 0.603). Top row: original T2-weighted images; bottom row: corresponding model predictions overlaid with ground truth annotations.

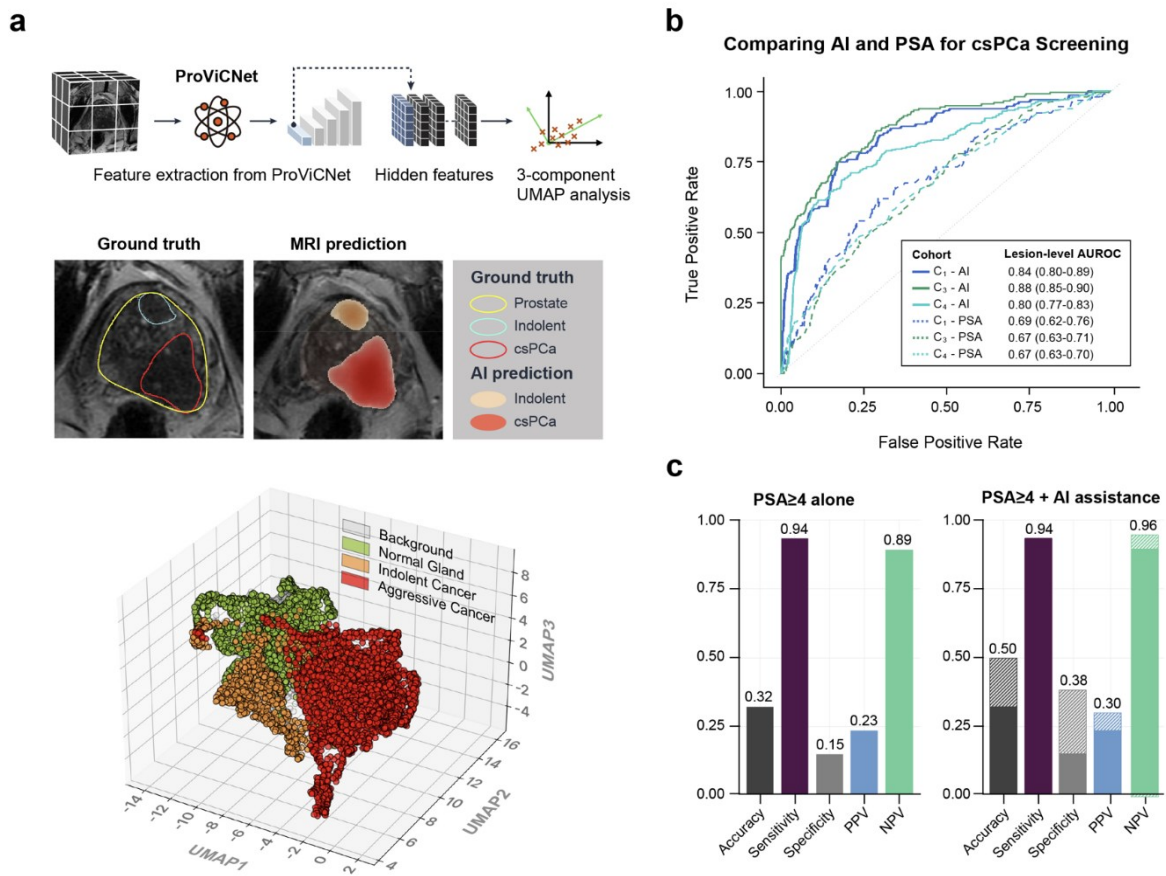

**Fig. 3 | ProViCNet feature representation analysis and application for screening to reduce unnecessary biopsies. a**, Visualization of learned feature representations using dimensionality reduction. ProViCNet features extracted from image patches of the T11 cohort were projected into 3D space using UMAP dimensional reduction, showing clear separation between normal prostate tissue (green), indolent cancer (orange), and aggressive cancer (red) regions. **b**, Receiver operating characteristic (ROC) curves comparing ProViCNet's performance (blue) against PSA screening (yellow) for detecting csPCa at patient level (AUROC: 0.798-0.875 vs 0.666-0.668, $p<0.001$). **c**, Performance metrics comparing PSA≥4 screening alone (left) versus PSA≥4 with AI assistance (right) for biopsy decision support. AI assistance maintained high sensitivity (0.937) while significantly reducing unnecessary biopsies through improved specificity (0.147 to 0.378, 157% increase).

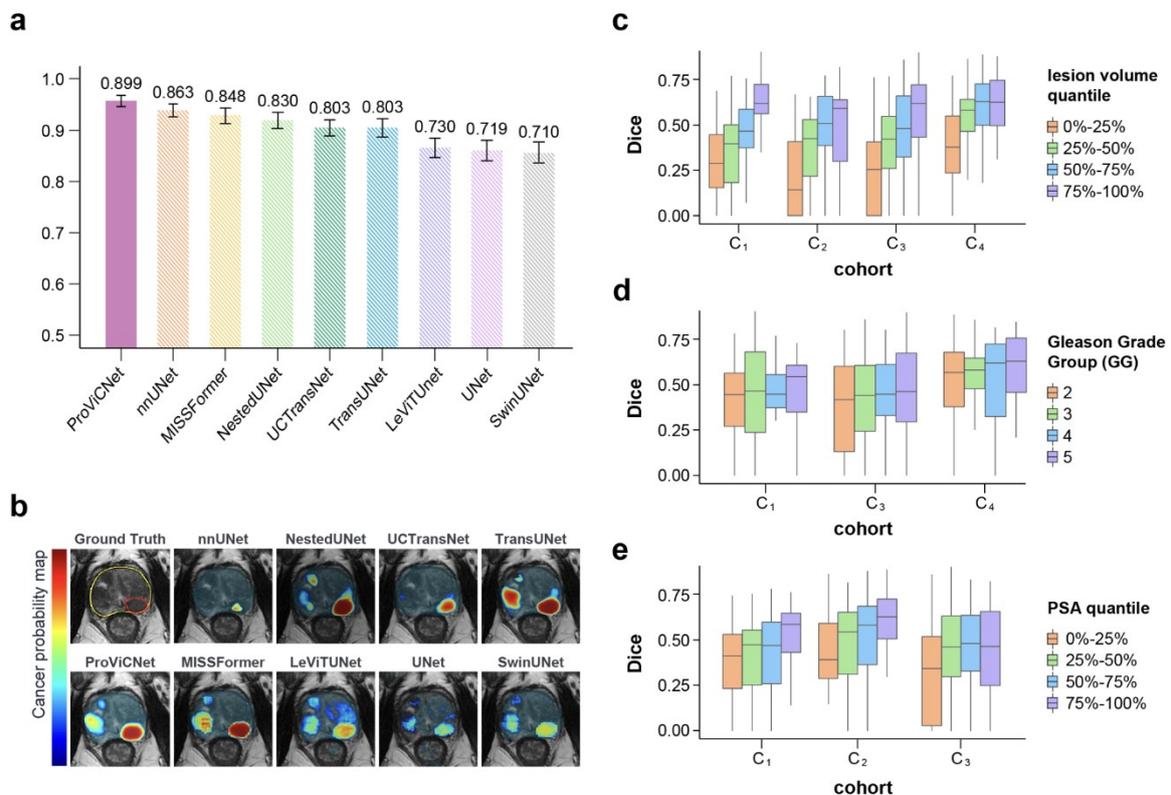

**Fig. 4 | Comparative analysis of model performance with other segmentation models and**

**its correlation with clinical variables. a**, Comparison of model performance against other established deep learning architectures. Bar plots show AUROC values with 95% confidence intervals for nine different models evaluated on T2-weighted MRI sequences from the T11 cohort. ProViCNet achieves superior performance (AUROC 0.899) compared to other architectures. **b**, Qualitative comparison of cancer probability predictions across different models on a representative case. Ground truth annotations (yellow outline) and predicted cancer probability maps are shown for each model, demonstrating ProViCNet's improved cancer localization accuracy. **c-e**, Clinical correlates of model performance across different cohorts: **c**, Dice scores stratified by lesion volume quartiles showing improved performance for larger lesions, **d**, Dice scores across different Gleason Grade Groups demonstrating consistent performance across cancer grades, and **e**, Dice scores stratified by PSA level quartiles indicating model performance is largely independent of PSA levels.

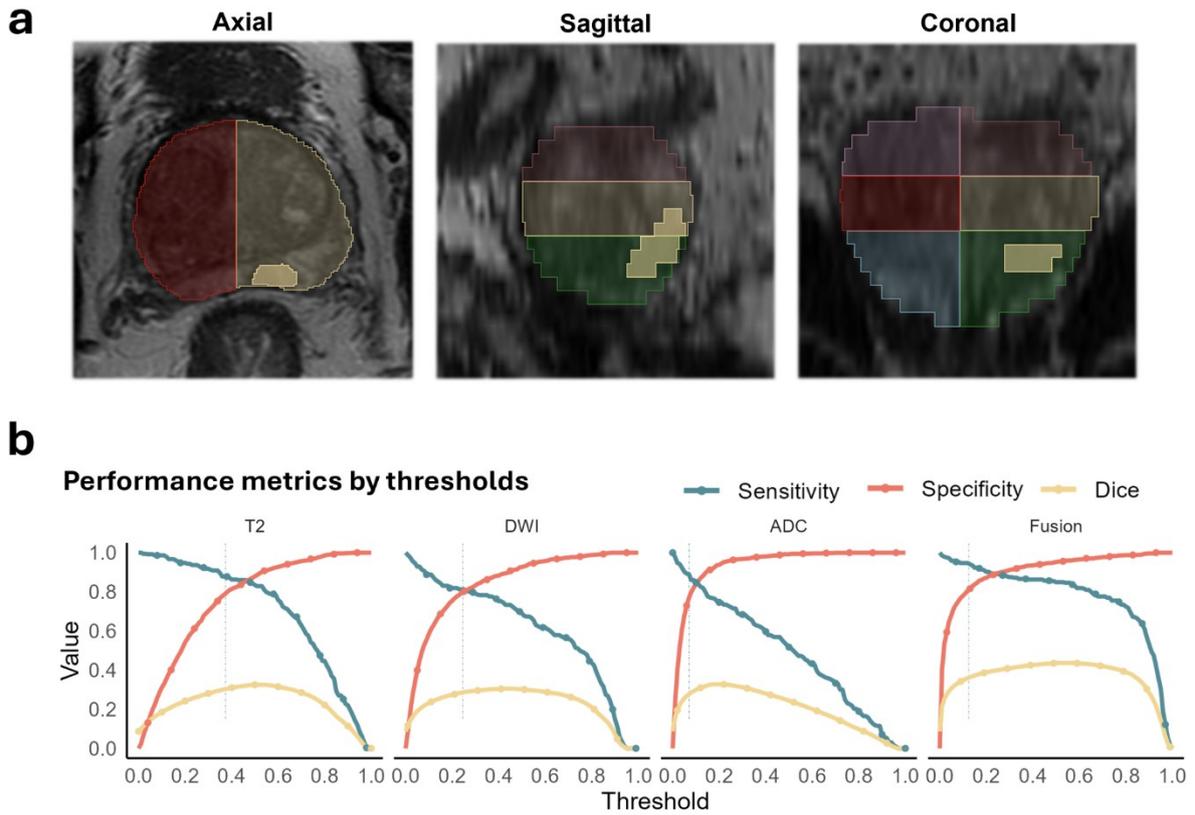

**Extended Data Fig. 1 | Lesion-level evaluation framework for prostate cancer detection.** **a**, Three-dimensional segmentation of prostate sextants shown in axial, sagittal, and coronal views. Color-coded regions represent the anatomical division used for region-based evaluation of cancer detection. **b**, Performance metrics (sensitivity, specificity, and Dice score) across different probability thresholds shown separately for T2-weighted, DWI, ADC sequences and their fusion. The optimal threshold values were determined from the development dataset.

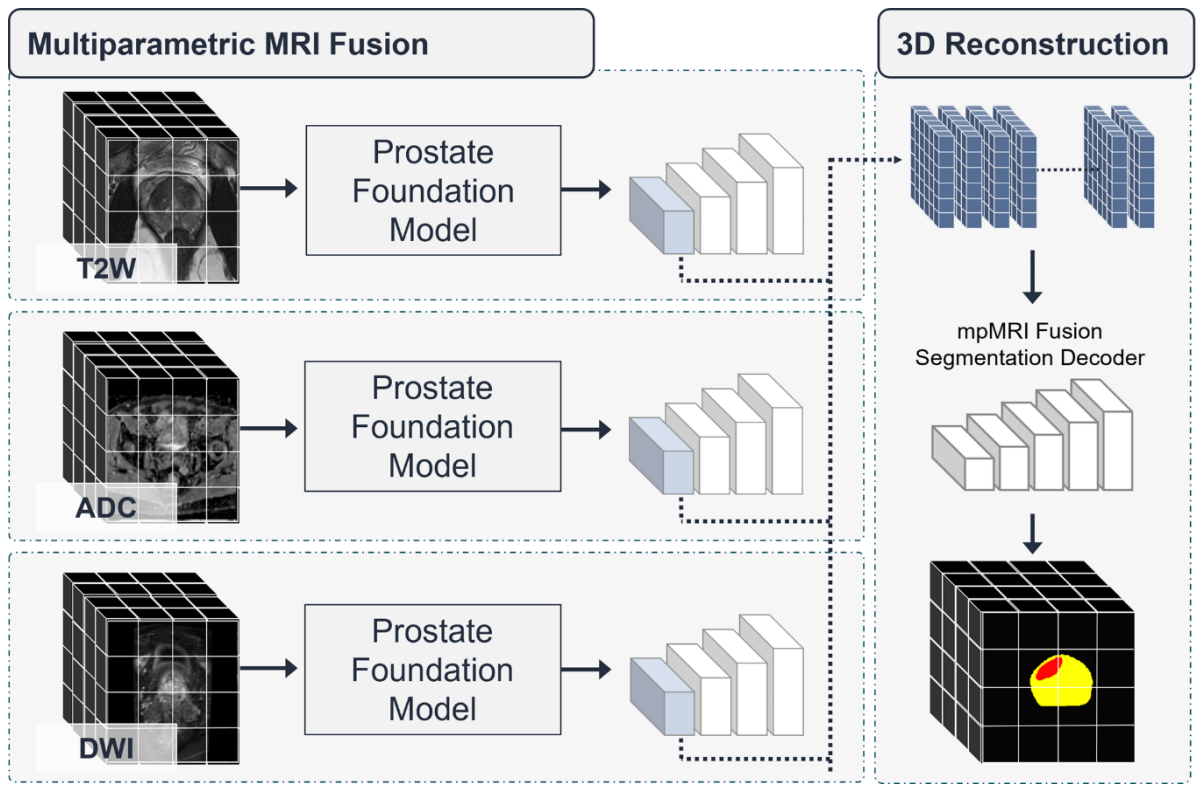

**Extended Data Fig. 2 | Multi-parametric MRI fusion architecture of ProViCNet.** Schematic overview of multi-sequence integration pipeline. T2W, ADC, and DWI sequences are independently processed through prostate foundation model encoders to extract sequence-specific features. These features are combined through a 3D reconstruction pathway and processed by an mpMRI fusion segmentation decoder to generate comprehensive cancer detection maps.

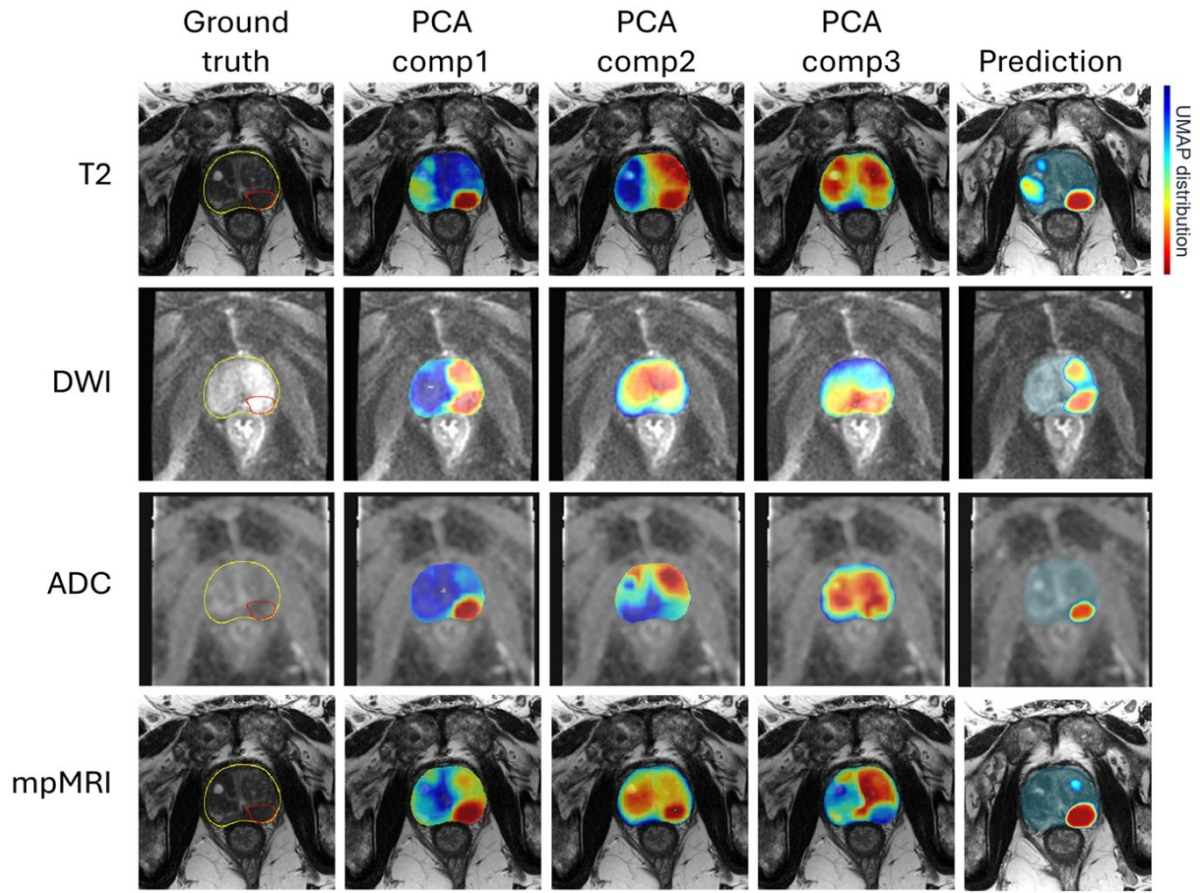

**Extended Data Fig. 3** | Feature visualization of multi-parametric MRI sequences. a, Overview of feature extraction and visualization pipeline. Small patches from prostate regions were processed through ProViCNet to extract features, which were then dimensionally reduced using PCA and interpolated back to original image dimensions. b, Visualization of the first three PCA components and final predictions across different MRI sequences (T2-weighted, DWI, ADC) and their multi-parametric fusion (mpMRI). Each column shows ground truth annotations, individual PCA components' spatial distributions, and the model's final prediction. The PCA components reveal distinct spatial patterns of feature organization within the prostate gland, with the final predictions integrating information across these components to localize cancer regions. Yellow outlines indicate prostate gland boundaries, and red regions denote annotated cancer areas in ground truth images.

**Table 1 | Patient and dataset characteristics across internal and external cohorts.**

| Description | | Internal Cohort | | | External Cohorts | | |
|---|---|---|---|---|---|---|---|
| | | Development Dataset | Test ($C_1$) | Test ($C_2$) | PICAI ($C_3$) | UCLA ($C_4$) | UCSD ($C_5$) |
| Source Procedure | | Biopsy | Biopsy | RP | Biopsy | Biopsy | Biopsy |
| Image modality | | MRI/TRUS | MRI/TRUS | MRI/ | MRI | MRI/TRUS | MRI |
| Population | | | | | | | |
| # Patients | | 1404 | 352 | 120 | 1473 | 760 | 292 |
| PSA Mean ± standard deviation [quantile 25% - 75%] | | 9.7 ± 14.0 [5.0-10.5] | 9.0 ± 6.9 [5.0-10.5] | 9.1 ± 5.3 [5.6-11.0] | 11.9 ± 15.0 [5.9-13.0] | 8.9 ± 10.5 [4.6-10.0] | 9.9 ± 12.6 [4.6-9.7] |
| Age Mean ± standard deviation [quantile 25% - 75%] | | 65.7 ± 7.4 [61.-71] | 65.2 ± 7.4 [60-71] | 63.6 ± 7.9 [60-69] | 65.6 ± 7.2 [61-70] | 65.5 ± 7.5 [61-71] | 68.8 ± 8.29 [62-74] |
| # Scans | | 1404 | 352 | 120 | 1497 | 1154 | 292 |
| # GG=0 #(%) | | 609 (42.3%) | 155 (44.0%) | 1 (0.8%) | 1091 (72.9%) | 387 (33.5%) | 139 (47.6%) |
| # GG=1 #(%) | | 245 (17.0%) | 43 (12.2%) | 23 (19.2%) | - | 286 (24.8%) | 56 (19.2%) |
| # GG≥2 #(%) | | 550 (40.7%) | 154 (43.8%) | 96 (80.0%) | 406 (27.1%) | 481 (41.7%) | 43 (14.7%) |
| Lesions | | | | | | | |
| # | Cancer | 1137 | 271 | - | 436 | 972 | - |
| # | Sign Cancer | 730 | 187 | - | 232 | 622 | 114 |
| # | GG=1 #(%) | 407 (35.8%) | 84 (31.0%) | - | 204 (46.8%) | 350 (36.0%) | - |
| # | GG=2 #(%) | 354 (31.1%) | 100 (36.9%) | - | 138 (31.7%) | 392 (40.3%) | 50 (43.9%) |
| # | GG=3 #(%) | 205 (18.0%) | 55 (20.3%) | - | 54 (12.4%) | 120 (12.3%) | 27 (23.7%) |
| # | GG=4 #(%) | 97 (8.5%) | 14 (5.2%) | - | 20 (4.6%) | 60 (6.2%) | 17 (14.9%) |
| # | GG=5 #(%) | 74 (6.5%) | 18 (6.6%) | - | 20 (4.6%) | 50 (5.1%) | 20 (17.5%) |
| MRI Characateristics | | | | | | | |
| Prostate size (range, mm) | | 56.7 ± 33.6 | 56.2 ± 34.2 | 28.7 ± 14.1 | 58.9 ± 31.6 | 48.5 ± 25.7 | 49.3 ± 38.3 |
| Cancer size (range, mm) | | 2.7 ± 13.6 | 2.4 ± 6.2 | 1.2 ± 3.4 | 1.51 ± 7.0 | 2.1 ± 6.8 | 1.1 ± 4.8 |
| Number of Axial Slices | | 33.2 ± 4.4 | 33.4 ± 4.5 | 35.5 ± 4.2 | 20.0 ± 0.0 | 30.0 ± 1.3 | 32.5 ± 1.1 |

**Table 2. Ablation study**

| Model Name | DINOv2 pretrained weights | Low-Rank Adaptation (LoRA) | Positional EmbedTokens (3D-Enhanced) | Patch-level Contrastive learning | AUROC | Sensitivity | Specificity |
|---|---|---|---|---|---|---|---|
| ViT | | | | | 0.747 | 0.635 | 0.758 |
| DINOv2 ViT | ✓ | | | | 0.877 | 0.763 | 0.903 |
| DINOv2 LoRA ViT | ✓ (frozen) | ✓ | | | 0.824 | 0.735 | 0.708 |
| ProViDNet | ✓ | | ✓ | | 0.918 | 0.883 | 0.702 |
| ProViCNet (Ours) | ✓ | | ✓ | ✓ | 0.930 | 0.872 | 0.808 |

**Extended Data Table 1 | ProViCNet performance metrics for clinically significant prostate cancer detection across cohorts.**

| Cohort | AUROC | AUPROC | Sensitivity | Specificity | PPV | NPV | DSC | Accuracy |
|---|---|---|---|---|---|---|---|---|
| Internal cohorts | | | | | | | | |
| $C_1$ | 0.923 | 0.879 | 0.895 | 0.778 | 0.444 | 0.991 | 0.425 | 0.793 |
| $C_2$ (RP) | 0.875 | 0.822 | 0.819 | 0.730 | 0.597 | 0.903 | 0.389 | 0.772 |
| External cohorts | | | | | | | | |
| $C_3$ | 0.966 | 0.933 | 0.953 | 0.761 | 0.360 | 0.998 | 0.526 | 0.767 |
| $C_4$ | 0.920 | 0.854 | 0.846 | 0.766 | 0.358 | 0.986 | 0.417 | 0.773 |
| $C_5$ | 0.973 | 0.964 | 0.850 | 0.951 | 0.604 | 0.988 | 0.451 | 0.943 |

**Extended Data Table 2 | Performance analysis of individual MRI sequences.**

| Sequence | AUROC | AUPROC | Sensitivity | Specificity | PPV | NPV | Dice | Accuracy |
|---|---|---|---|---|---|---|---|---|
| *Dataset: $T_{11}$ Clinically significant lesions only* | | | | | | | | |
| T2 | 0.899 | 0.833 | 0.774 | 0.874 | 0.546 | 0.981 | 0.320 | 0.868 |
| ADC | 0.885 | 0.830 | 0.741 | 0.885 | 0.542 | 0.976 | 0.301 | 0.875 |
| DWI | 0.851 | 0.773 | 0.816 | 0.699 | 0.311 | 0.961 | 0.235 | 0.721 |
| *Dataset: $T_{12}$ Clinically significant lesions only* | | | | | | | | |
| T2 | 0.824 | 0.775 | 0.817 | 0.636 | 0.551 | 0.896 | 0.308 | 0.724 |
| ADC | 0.866 | 0.850 | 0.569 | 0.948 | 0.839 | 0.808 | 0.197 | 0.799 |
| DWI | 0.827 | 0.765 | 0.716 | 0.733 | 0.603 | 0.850 | 0.270 | 0.732 |

**Extended Data Table 3 | ProViCNet performance for transrectal ultrasound (TRUS)**

imaging.

| Cohort | AUROC | AUPROC | Sensitivity | Specificity | PPV | NPV | Dice | Accuracy |
|---|---|---|---|---|---|---|---|---|
| $C_1$ | 0.735 | 0.596 | 0.691 | 0.571 | 0.245 | 0.958 | 0.144 | 0.584 |
| $C_4$ | 0.670 | 0.539 | 0.715 | 0.462 | 0.205 | 0.949 | 0.122 | 0.476 |

**Extended Data Table 4 | ProViCNet performance for all prostate cancers including indolent disease.**

| Cohort | AUROC | AUPROC | Sensitivity | Specificity | PPV | NPV | Dice | Accuracy |
|---|---|---|---|---|---|---|---|---|
| Internal cohorts | | | | | | | | |
| $C_1$ | 0.942 | 0.914 | 0.942 | 0.647 | 0.357 | 0.994 | 0.409 | 0.675 |
| $C_2$ | 0.907 | 0.861 | 0.885 | 0.628 | 0.544 | 0.923 | 0.394 | 0.724 |
| External cohorts | | | | | | | | |
| $C_4$ | 0.925 | 0.871 | 0.922 | 0.636 | 0.286 | 0.992 | 0.410 | 0.659 |

**Extended Data Table 5 | Comparison of ProViCNet and expert reader performance on 91 radical prostatectomy cases from cohort T12, evaluated at patient and lesion levels.**

| Reader | AUROC | AUPROC | Sensitivity | Specificity | PPV | NPV | Dice | Accuracy |
|---|---|---|---|---|---|---|---|---|
| Patient-level evaluation | | | | | | | | |
| AI | 0.907 | 0.861 | 0.880 | 0.654 | 0.561 | 0.924 | 0.396 | 0.741 |
| Reader | 0.805 | 0.738 | 0.825 | 0.799 | 0.648 | 0.892 | 0.347 | 0.800 |
| Lesion-level evaluation | | | | | | | | |
| AI | 0.867 | N/A | 0.866 | 0.661 | 0.448 | 0.939 | N/A | 0.710 |
| Reader | 0.771 | N/A | 0.787 | 0.787 | 0.556 | 0.916 | N/A | 0.787 |

**Extended Data Table 6 | Comparative analysis of deep learning segmentation models.**

| Reader | AUROC | AUPROC | Sensitivity | Specificity | PPV | NPV | Dice | Accuracy |
|---|---|---|---|---|---|---|---|---|
| SwinUNet | 0.710 | 0.607 | 0.605 | 0.640 | 0.238 | 0.925 | 0.149 | 0.631 |
| UNet | 0.719 | 0.667 | 0.173 | 0.977 | 0.717 | 0.894 | 0.068 | 0.879 |
| LeViTUnet | 0.730 | 0.631 | 0.220 | 0.914 | 0.353 | 0.895 | 0.050 | 0.824 |
| TransUNet | 0.803 | 0.737 | 0.846 | 0.496 | 0.212 | 0.961 | 0.189 | 0.542 |
| UCTransNet | 0.803 | 0.711 | 0.621 | 0.824 | 0.374 | 0.937 | 0.207 | 0.798 |
| NestedUNet | 0.830 | 0.749 | 0.828 | 0.666 | 0.288 | 0.961 | 0.222 | 0.694 |
| MISSFormer | 0.848 | 0.802 | 0.872 | 0.536 | 0.225 | 0.958 | 0.208 | 0.584 |
| nnUNetT22D | 0.869 | 0.795 | 0.256 | 0.996 | 0.868 | 0.932 | 0.105 | 0.929 |
| ProViCNet (ours) | 0.899 | 0.833 | 0.774 | 0.874 | 0.546 | 0.981 | 0.320 | 0.868 |

**Extended Data Table 7 | Performance comparison of nnUNet and ProViCNet.**

| AI | AUROC | AUPROC | Sensitivity | Specificity | PPV | NPV | Dice | Accuracy |
|---|---|---|---|---|---|---|---|---|
| *Dataset: $C_{11}$ Clinically significant lesions only* | | | | | | | | |
| ProViCNet | 0.899 | 0.833 | 0.774 | 0.874 | 0.546 | 0.981 | 0.320 | 0.868 |
| nnUNet | 0.863 | 0.778 | 0.476 | 0.974 | 0.732 | 0.952 | 0.273 | 0.909 |
| *Dataset: $C_{11}$ Including indolent cancer* | | | | | | | | |
| ProViCNet | 0.893 | 0.823 | 0.934 | 0.426 | 0.229 | 0.988 | 0.329 | 0.478 |
| nnUNet | 0.869 | 0.795 | 0.664 | 0.905 | 0.544 | 0.972 | 0.305 | 0.891 |

**Extended Data Table 8 | Distribution of MRI scanner manufacturers in internal cohort.**

| Manufacturer | Number of cases | Percentage (%) |
|---|---|---|
| GE MEDICAL SYSTEMS | 1538 | 84.55% |
| Philips Medical Systems | 189 | 10.39% |
| SIEMENS | 92 | 5.06% |

**Extended Data Table 9 | Checklist for supervised clinical ML study**

| Before paper submission | | | |
|---|---|---|---|
| **Study design (Part 1)** | | **Completed: page number** | **Notes if not completed** |
| The clinical problem in which the model will beemployed is clearly detailed in the paper. | ☒ | Page 4 | |
| The research question is clearly stated. | ☒ | Page 4 | |
| The characteristics of the cohorts (training | ☒ | Table 1, Figure 1c | |
| andtest sets) are detailed in the text. | ☒ | | |
| The cohorts (training and test sets) are shownto be representative of real-world clinical settings. | ☒ | Page 5 | |
| The state-of-the-art solution used as a baselinefor comparison has been identified and detailed. | ☒ | Figure 4 Extended Data Table 6 | |
| **Data and optimization (Parts 2, 3)** | | **Completed: page number** | **Notes if not completed** |
| The origin of the data is described and theoriginal format is detailed in the paper. | ☒ | Page 19 | |
| Transformations of the data before it is appliedto the proposed model are described. | ☒ | Page 22 | |
| The independence between training and testsets has been proven in the paper. | ☒ | Table 1 | |
| Details on the models that were evaluated andthe code developed to select the best model are provided. | ☒ | Page 28 | |
| Is the input data type structured orunstructured? | | ☒ Structured ☐ Unstructured | |
| **Model performance (Part 4)** | | **Completed: page number** | **Notes if not completed** |
| The primary metric selected to evaluate algorithm performance (eg: AUC, F-score, etc)including the justification for selection, has been clearly stated. | ☒ | Page 26 | |
| The primary metric selected to evaluate the clinical utility of the model (eg PPV, NNT, etc)including the justification for selection, has been clearly stated. | ☒ | Page 26 | |
| The performance comparison between baseline and proposed model is presented withthe appropriate statistical significance. | ☒ | Page 12 Extended data table 6 | |
| **Model Examination (Parts 5)** | | **Completed: page number** | **Notes if not completed** |
| Examination Technique 1[a]: Model performance over time | ☒ | 3, 4, Figs. 3, 5 | |
| Examination Technique 2[a]: SHAP value analysis | ☐ | | Not applicable as the model uses image data |
| A discussion of the relevance of the examination results with respect to | ☒ | Page 16 | |
| A discussion of the feasibility and significanceof model interpretability at the case level if examination methods are uninterpretable is presented. | ☒ | Page 10 | UMAP feature representation analysis |
| A discussion of the reliability and robustness ofthe model as the underlying data distribution shifts is included. | ☒ | Page 17. | multiple external validation cohorts |

| *Common examination approaches based on study type: <br> * For studies involving exclusively structured data coefficients and sensitivity analysis are often appropriate <br> * For studies involving unstructured data in the domains of image analysis or NLP: saliency maps (or equivalents) and sensitivity analysis are often appropriate | ☒ | Page 10 | UMAP feature representation analysis |
|---|---|---|---|
| **Reproducibility (Part 6): choose appropriate tier of transparency** | | | **Notes** |
| Tier 1: complete sharing of the code | ☒ | | Both the code and model for ProViCNet will be publicly available at GitHub (https://github.com/pimed/ProViCNet) and Hugging Face (https://huggingface.co/pimed/ProViCNet). |
| Tier 2: allow a third party to evaluate the code for accuracy/fairness; share the results of this evaluation | ☐ | | |
| Tier 3: release of a virtual machine (binary) for running the code on new data without sharing its details | ☐ | | |
| Tier 4: no sharing | ☐ | | |